\documentstyle[12pt,aaspp4]{article} 
\newcommand{\kms}          {\mbox{${\rm km~s^{-1}}$}}
\newcommand{\cc}           {\mbox{${\rm cm^{-3}}$}}

\newcommand{\simgt}        {\gtrsim}

\newcommand{\Msun}         {\mbox{$\rm M_\odot$}}

\newcommand{\HI}           {\mbox{${\rm H{\scriptstyle I}}$}}
\newcommand{\HII}          {\mbox{${\rm H{\scriptstyle II}}$}}

\newcommand{\plm}	   {\mbox{$\pm$}}
\def\l{$l$}
\def\b{$b$}
\def\amine{$^{\prime}$\ }
\def\cm2{\mbox{${\rm cm^{-2}}$}}
\def\deg{\ifmmode^\circ\else$^\circ$\fi}
\def\dege{{\ifmmode^\circ\else$^\circ$\fi}\ }
\def\degper{\ifmmode \rlap.^{\circ} \else $\rlap{.}^{\circ} $\fi}

\def\2{$^{2}$}
\begin{document}

\title{Gas Rich Dwarf Spheroidals }

\author{Leo Blitz}
\author{Timothy Robishaw}
\vspace{.4in}
\affil{Astronomy Department, University of California, Berkeley, CA
94720;\\
blitz@gmc.berkeley.edu; robishaw@gmc.berkeley.edu}


\begin{abstract}
\rightskip = 0pt
\noindent
 	We present evidence that nearly half of the dwarf spheroidal 
galaxies (dSph and dSph/dIrr) in the Local Group are associated with 
large reservoirs of
atomic gas, in some cases larger than the stellar mass.  The gas is
sometimes found at large distance ($\sim$10 kpc) from the center of a
galaxy and is not necessarily centered on it.  Similarly large
quantities of ionized gas could be hidden in these systems as well.
The properties of some of the gas reservoirs  are similar to the median
properties of the High-Velocity Clouds (HVCs); two of the \HI\ reservoirs
are catalogued HVCs.  The association of the \HI\ with the dwarf
spheroidals might thus provide a link between the 
HVCs and stars.  We show that the \HI\ content of the Local Group
dSphs and dIrrs exhibits a sharp decline if the galaxy is within 250 kpc
of either the Milky Way or M31.  This can be explained if both galaxies
have a sufficiently massive x-ray emitting halo that produces
ram-pressure stripping if a dwarf ventures too close to either giant
spiral. We also investigate tidal stripping of the dwarf galaxies and
find that although it may play a role, it cannot explain the apparent
total absence of neutral gas in most dSph galaxies at distances less
than 250 kpc. For the derived mean density of the hot gas, $n_{\circ} =
2.5 \times 10^{-5}$ \cc, ram-pressure stripping is found to be more
than an order of magnitude more effective in removing the gas from the
dSph galaxies.  The hot halo, with an inferred mass of $1 \times 10^{10}$
\Msun, may represent a reservoir of $\sim 10^3$ destroyed dwarf systems, 
either HVCs or true dwarf galaxies similar to those we observe now. 
\end{abstract}


\rightskip = 0pt
\section{Introduction}

Recently, Blitz et al.\ (1999) have shown that the HVCs, clouds of
atomic hydrogen inconsistent with near-circular rotation about the
Galactic Center, are well explained if they are members of the Local
Group.  A simple dynamical model can replicate both the observed
distribution on the sky as well as the observed kinematics of the
ensemble of  the HVCs.  In their paper, Blitz et al.\ (1999) made
several predictions; among them are that HVCs have  
H$\alpha$ surface brightnesses 
less than those measured in the Magellanic Stream, metallicities of
0.1 solar or less, and internal pressures P/k $\sim$ 1-10 K \cc;  the
last of these is inferred from the self-gravity of HVCs.  All of these
predictions have been subsequently confirmed (Weiner et al.\ 2000;
Wakker et al.\ 1999; Sembach et al.\ 1999).  Only one set of
observations now seems to present problems for the model: three deep,
blind extragalactic \HI\ surveys, two made at Arecibo (Zwaan et al.\ 1997;
Spitzak \& Schneider 1998), and one at Parkes (Banks et al.\ 1999).

If the HVCs are extragalactic and part of the Local Group, as Blitz et
al.\ (1999) argue, then extragalactic analogues should be observable;
many examples have indeed been reported in the literature (Blitz et
al.\ 1999 and references therein).  The blind
\HI\ surveys have also found many uncatalogued \HI\ clouds, but on
closer inspection, all but one of the detections are found to harbor
galaxies, albeit often of very low surface brightness.
Because stars have not been detected in HVCs, they have been assumed to
be starless systems.  No such
systems have been found by either Zwaan et al.\ (1997) or Banks et al.\ (1999);  and only one potentially
starless system has been found in the Spitzak \& Schneider (1998)
survey.  The sensitivity of both surveys appears to be high enough and
the velocity coverage large enough, that at least a few HVC analogues
should have been detected in these surveys.

On the other hand, it is unclear whether the lowest surface brightness
galaxies found in the optical follow-up to the \HI\ detections would have
been detected if those systems were located in the Local Group.  These
galaxies, which compose a significant fraction of the Zwaan et al.\ (1997)
and Spitzak \& Schneider (1998) \HI\ identifications, appear
morphologically  similar to the Local Group dwarf spheroidal galaxies,
which have been observed to be generally gas-free (Mateo 1998).  At
Local Group distances such low surface brightness systems would be
extended and relatively difficult to identify in existing
surveys.  However, with concerted searching, a number of dSph galaxies
have recently been discovered in the Local Group (Whiting et al.\ 1997;
Armandroff, Davies, \& Jacoby 1998; Karachentsev \& Karachentseva 1999;
Gallart et al.\ 1999), suggesting that there might be similar, or even
lower surface brightness galaxies associated with HVCs.  

A search for low surface brightness galaxies in HVCs is currently
underway, but as a first step, we decided to examine the newly
discovered Local Group dwarfs for 21-cm emission using the 
Leiden-Dwingeloo \HI\ survey (LDS - Hartmann \& Burton 1997) to see if
any are associated with HVCs.  After the detection of \HI\ toward And
V, and its subsequent identification as HVC 368 in the compilation of
Wakker \& van Woerden (1991), 
we decided to reexamine the \HI\ content of all of the dSph galaxies
in the Local Group; the results are presented in \S 3.
We have found \HI\ toward four galaxies in which it had not previously
been detected, and found \HI\ more extended than previously thought in two
others.  The \HI\ found toward two galaxies are catalogued HVCs.
In \S 4 we examine the implications of the \HI\ observations
and infer the existence of a hot gaseous corona around the Milky Way and
M31 with a mean density of $\sim 2.5 \times 10^{-5}$ \cc.  Rather than
being gas-poor, the dSph galaxies are often gas-rich, but with rather
extended \HI\ envelopes.

\section{Analysis}
\label{sec:analysis}

The sensitivity of the LDS is about 70 mK in a 1 \kms\ velocity
channel; its angular resolution is 36\arcmin.  The survey covers the
entire northern sky down to a declination of -30\dege at a sampling
interval of 30\arcmin. At a distance of 100 kpc, the beam is 1.05 kpc.
Its effective velocity coverage of -450 $ \rm \leq V_{LSR} \leq$ +400
\kms\ is sufficient to detect all Local Group emission down to a
5$\sigma$ column density N(H)/$(\Delta V)^{-1/2} = 6.4 \times 10^{17}$
\cm2 (\kms)$^{-1/2}$ averaged over the beam ($\Delta V$ is the full width
at half maximum of the \HI\ emission).  The survey also has the
virtue of having flat baselines which makes it possible to detect very
low level emission at velocities away from the normal Galactic
emission.

We initially examined a five point cross centered at the optical
position of all of the galaxies catalogued as dSph or dSph/dIrr in the
compilation of Mateo (1998), but excluded those also classified as dE or E
systems because of their much higher surface brightness.  We also
examined the galaxies subsequently identified as possible Local Group
dSph galaxies by Karachentsev \& Karachentseva (1999) and Gallart et
al.\ (1999) which could be found in the LDS. If a galaxy sits between
grid points of the survey, we examined a somewhat larger area.  A
number of the galaxies were found to have 21-cm emission 
confined to a small area coincident with or very close to the dSph
at velocities outside the range normally associated with Galactic emission. 

One can estimate the probability of a chance coincidence of a dSph with
a cloud along the line of sight from the surface filling fraction of
small HVCs ($\leq$ 1 deg\2) in the Wakker \& van Woerden (1991)
compilation.  The HVCs comprise all of the \HI\ emission not associated
with galaxies outside the range of normal Galactic emission and have
the same range of radial velocities as the dSphs.  The total area on
the sky of small HVCs is less than 300 deg\2, though this number is
somewhat uncertain at the 50\% level because of the sparse sampling of
the Wakker \& van Woerden (1991) catalogue.  It is certainly an upper
limit, though, based on the higher resolution mapping of about 20\% of
the smaller HVCs by Blitz et al.\ (1999) and by Braun \& Burton
(1999).  The HVCs are spread all over the sky with some concentration
in the general directions of the barycenter and antibarycenter of the
Local Group, thus the probability of a chance coincidence with a cloud
that subtends an angle of less than 1 deg\2 is $\la 0.01$.
Equivalently, a positional coincidence with such a cloud is significant
at the $\ga 2.5 \sigma$ level.  We therefore consider the likelihood of a chance coincidence of a dSph with a small \HI\ cloud to be sufficiently low
that it is indicative of a real physical association, even if no
velocity is available for the galaxy.  Indeed, in a sample of $\sim$20
objects such as considered here, the probability of a chance positional
coincidence of {\it any} dSph with an \HI\ cloud is $\sim$0.2, and the
number of detections is about 50 times higher.

In the case where velocity information is available, the probability of
an optical and \HI\ velocity coincidence within 2$\sigma$ of the \HI\
velocity dispersion of $\sim$13 \kms\ (Blitz et
al.\ 1999) is about 0.06 within the velocity range in which HVCs are detected:
about 800 \kms.  This probability must be increased somewhat
because we consider only velocities outside the range of normal
Galactic emission, but in the directions in which most dSphs are found,
the Galactic emission is not wider than about 200 \kms. The joint
probability of velocity and spatial coincidence is thus $\sim 
10^{-3}$.  Probabilities of chance coincidences can be estimated for
individual galaxies and are done so where appropriate in the text. 

If emission was found to be associated with a galaxy, based on
positional coincidence alone, or where possible, joint position and velocity
coincidence, we averaged all of the profiles where emission was evident.
One or more Gaussians were then fit to the averaged spectrum in order to
determine a central velocity and velocity extent of the emission.  The
maps are shown in Figure 1. The values of the central velocity, FWHM
velocity extent, and brightness temperature of the Gaussian fits to the
position-averaged spectra are listed in Table 1.  The derived values
for M(${\rm H{\scriptstyle I}}$) and $\Omega$, the \HI\ mass and the
solid angle of the cloud, are not corrected for source convolution with
the telescope beam. We then searched a 7\deg$\times$7\deg\ area to
see if the emission is localized around the target galaxy or whether it
exists over a more extended area.  In two cases, rather extended
emission was found and the maps are shown in Figures 2 and 3. The
individual maps are discussed in \S\ref{sec:galaxies} below.

Some of the apparent detections were quite weak and Jay Lockman kindly
observed some of these with the NRAO 140\amine telescope\footnotemark
\footnotetext{The National Radio Astronomy Observatory is operated by
Associated Universities, Inc., under cooperative agreement with the
National Science Foundation.}
prior to its
shutdown in late July 1999 to obtain confirmation.  In most cases the
observations were about 4 times longer than those of the LDS.  The beam
of the 140\amine telescope is about 20\amine at the frequency of the 
21-cm line.

Finally, we searched the catalogues of Hulsbosch \& Wakker (1988) and
Wakker \& van Woerden (1991) to see whether any of the \HI\ clouds are
catalogued HVCs.  We found two such cases, the cloud associated with
And V is HVC 368, and the cloud associated with Sculptor is HVC 561.
LGS 3 had been  previously 
detected at three positions and DDO 210 at one position on
the 1\deg$\times$1\deg\ sampling grid of Hulsbosch \& Wakker (1988). 
The temperature-weighted mean Galactic latitude and longitude, as well as
the \HI\ column density and mass for each cloud can be found in Table
1.  The derived \HI\ masses assume that the emission is at the distance
of the galaxy. 
The non-detections are listed in Table 2.  The observed properties of 
the detected clouds are similar to one another
and to the typical properties of HVCs, but are generally weaker than
\HI\ detections of Local Group dIrr, Irr, or spiral galaxies.

\begin{figure}[htpb]
\plotfiddle{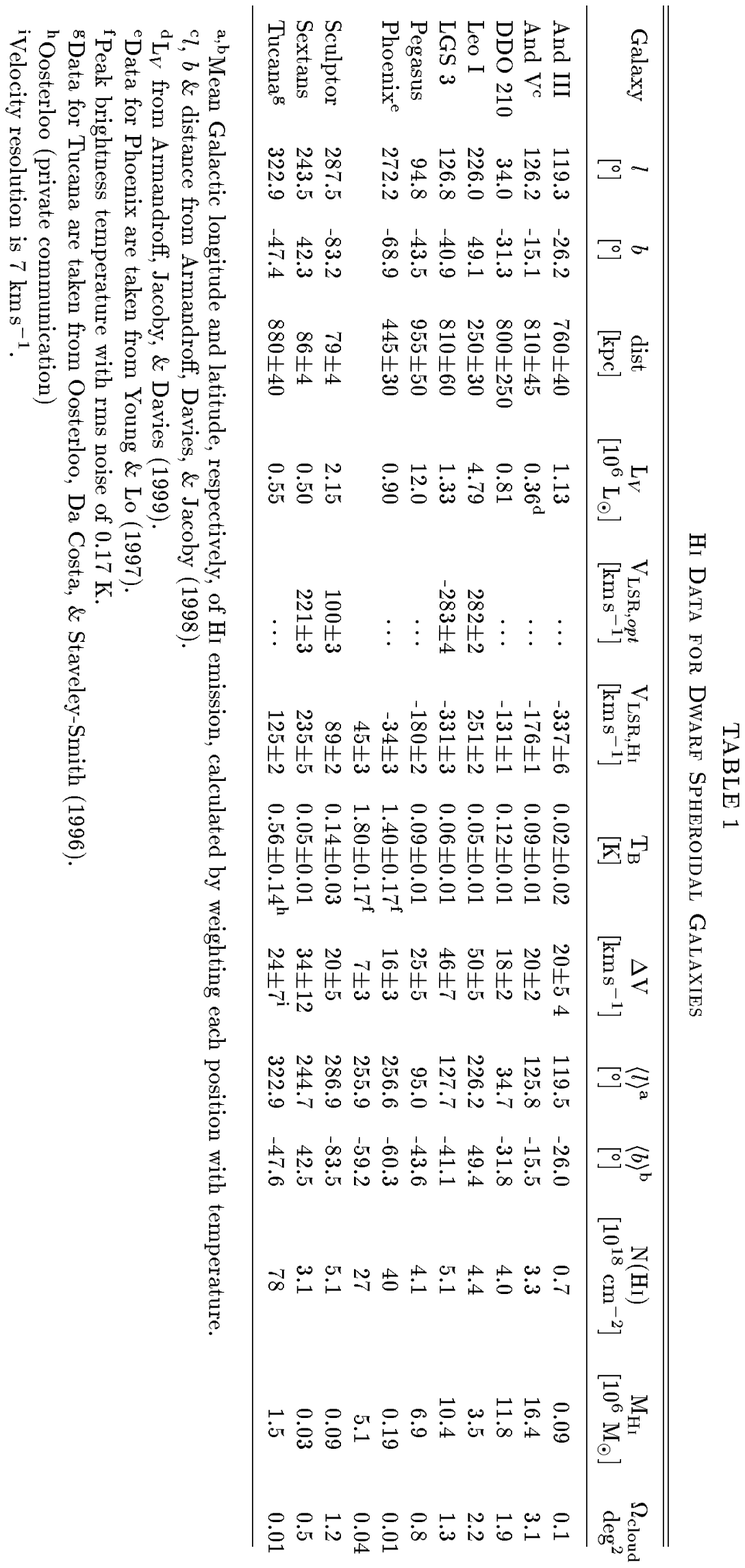}{160truemm}{180}{90}{90}{220}{575}
\end{figure}

\section{Results}
\label{sec:results}

We have made new \HI\ detections toward And III, And V, Leo I and Sextans.  In
addition, the \HI\ emission toward LGS 3, DDO 210, and Sculptor is found
to be considerably more extended than previously thought.  

In all, of 21 confirmed Local Group dSph galaxies, \HI\ is detected
in apparent association with 10 galaxies.  Two galaxies are too far
south to have been observed in the Leiden-Dwingeloo survey, and several
probable non-detections are somewhat ambiguous.  One galaxy previously
reported as a detection, Antlia (Fouqu\'e et al.\ 1990), is probably an
instrumental artifact.
There are 9 non-detections, not including the dE systems or the two
unconfirmed candidate dSph galaxies of Karachentsev \& Karachentseva
(1999).  Four galaxies are below the declination
limit of the LDS, but two of these, Phoenix and Tucana, have been
observed and detected in \HI\ by others (Young \& Lo 1997; Oosterloo et
al.\ 1996); they are therefore included in the list of detections.
Thus almost half of the confirmed Local Group dSph galaxies have been
detected in \HI, though only five of those detected have measured
optical velocities as of this writing.  Attempts are actively underway
by several groups to obtain more velocities.

Two of the galaxies shown in Figure 1 exhibit emission close
to, but not coincident with the target galaxy: Leo I and Sextans. 
This might explain why these two galaxies
have not been previously detected in \HI;  and why the detection 
of \HI\ toward Tucana by Oosterloo, Da Costa, \& Stavely-Smith (1996)
ought to be reinterpreted.  These authors found a cloud very close to,
but not
coincident with Tucana, in which the highest column density emission
is offset by about 15\arcmin~from the nucleus.  Oosterloo et al.\ (1996) 
felt that the position offset
implied that the \HI\ cloud and the galaxy are unrelated.  We show in the
next section, however, that the relationship between the \HI\ cloud and
Tucana is similar to other galaxies in our sample, particularly Leo I,
in which the velocity of the \HI\ and the galaxy are both measured and
found to be in close agreement.  We therefore include Tucana in Table 1
as a detection.

\begin{figure}[htpb]
\figurenum{1}
\plottwo{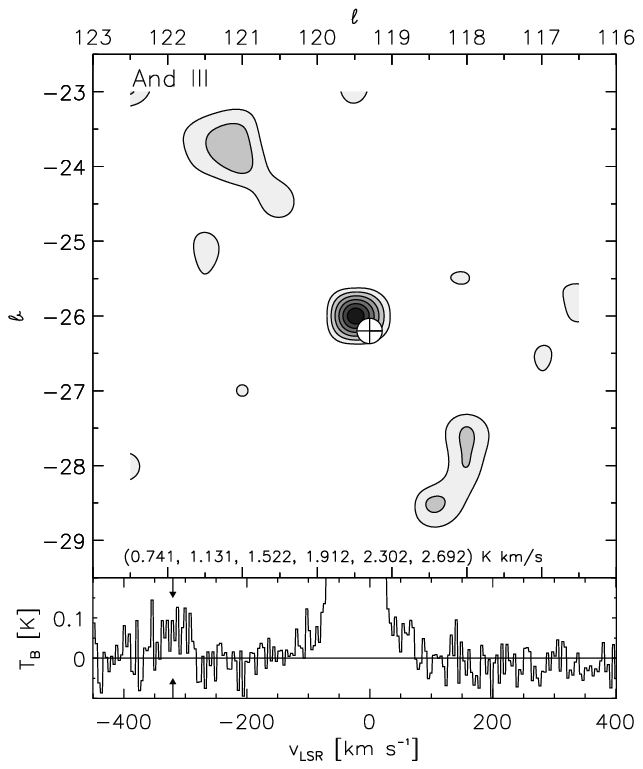}{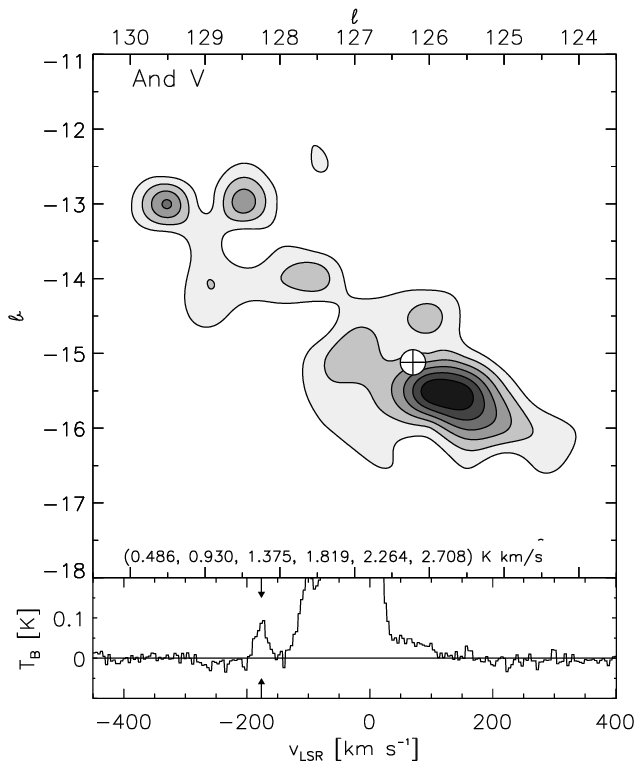} \vspace{.1in}
\epsscale{.45}
\plottwo{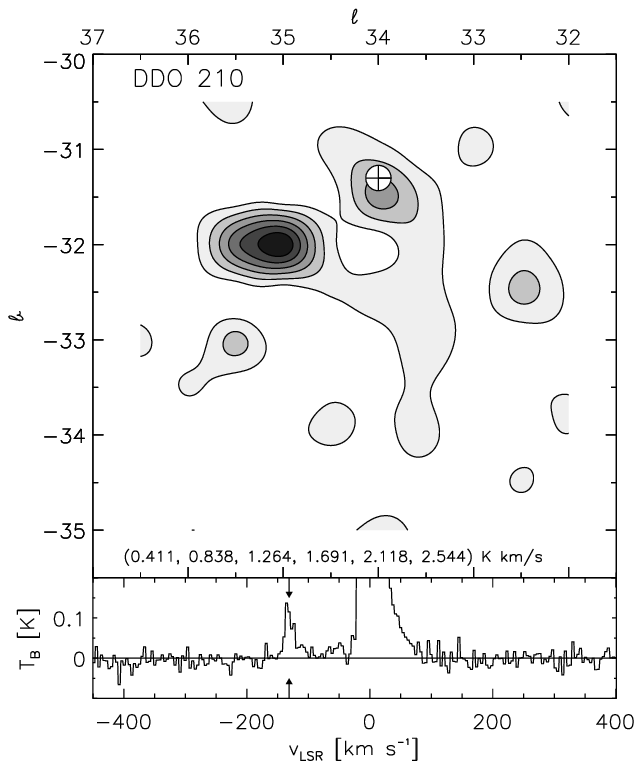}{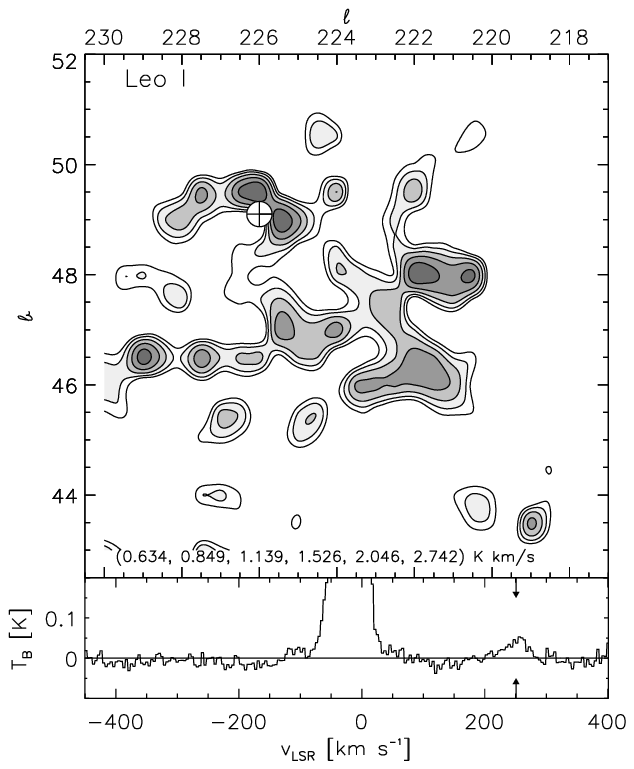}
\caption{
\baselineskip=15pt
Maps of the velocity-integrated \HI\ emission associated with Local
Group dSphs from the Leiden-Dwingeloo survey.  Below each map is a
spectrum of the emission averaged within the lowest contour of each
map. The position of the galaxy is given by the crossed circle.  The
lowest contour is 1.5 times the uncertainty of the velocity integrated brightness
temperature; the contour interval is 1/6 of the difference between the
first contour and the maximum value of the map.  The spectra have been
smoothed to 4.1 km/s resolution for easier viewing.  The arrow gives
the velocity centroid of the spectrum in each panel.
}
\end{figure}

\begin{figure}[htpb]
\figurenum{1 ({\it cont.})}
\plottwo{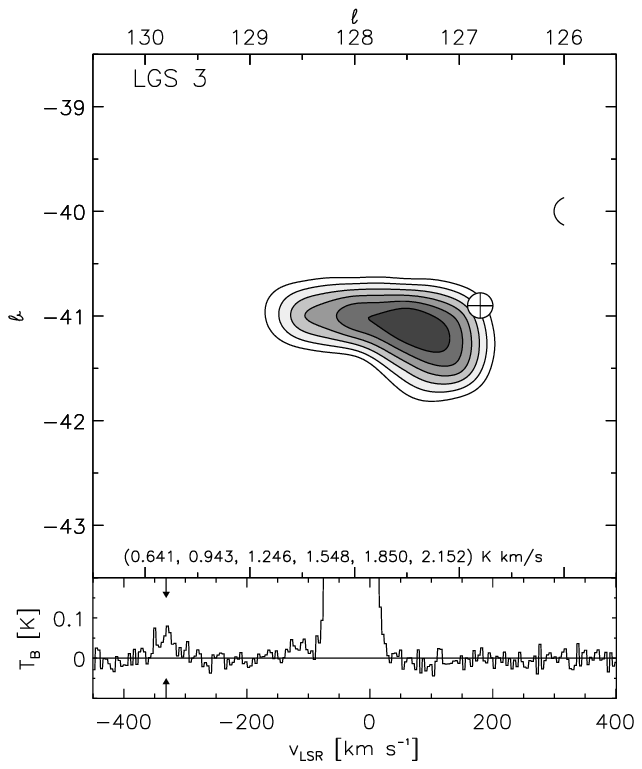}{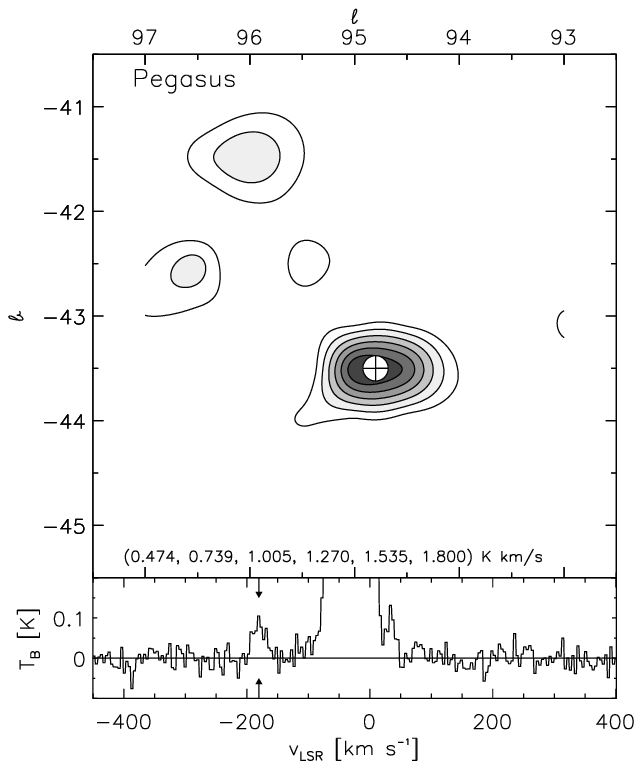} \vspace{.1in}
\epsscale{.45}
\plottwo{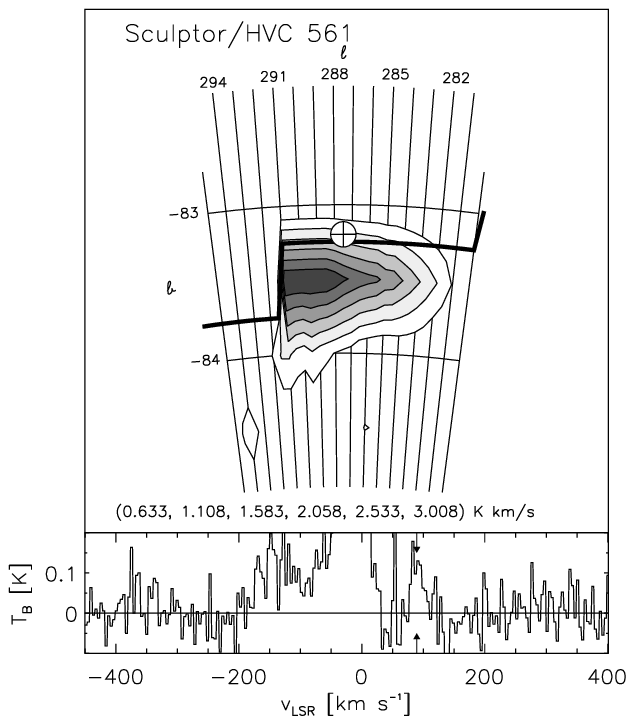}{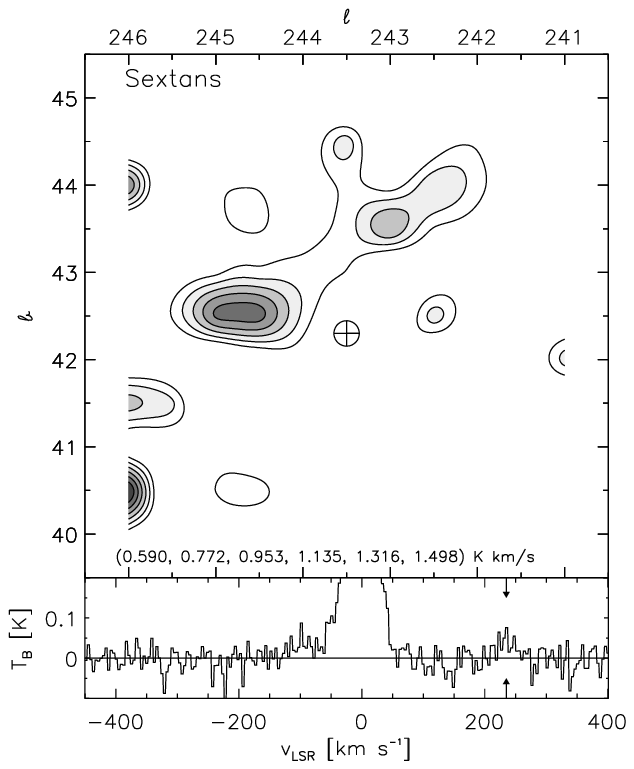}
\caption{The dark line shown for Sculptor is the lower declination
limit of the LDS.  A complete map is shown in Carignan (1999).}
\end{figure}

\subsection{Comments on Individual Detections}
\label{sec:galaxies}

\subsubsection {And III}

And III is a rather problematic \HI\ detection.  It
appears to be clearly detected in the LDS (see Figure 1), and a
confirmation spectrum of the galaxy taken with the 140\amine telescope
showed a much weaker component at low significance at the same
velocity.  The weakness of the 140\amine spectrum is rather surprising,
however, given the smaller beam and the longer integration time. 
The detection, nevertheless, does appear to be real, though of low
significance.  A measurement of the optical velocity of the galaxy or
a deep \HI\ integration over the LDS beam will confirm or refute this
detection.  The values given in Table 1 are from the deeper 140\amine
observations.

(NB: The referee, Mario Mateo, has pointed out that the optical
velocity of this galaxy has recently been determined by C\^ot\'e, Mateo,
\& Sargent (2000), and is within 1$\sigma$ of the combined \HI\ and
optical velocity uncertainties).

The \HI\ non-detection reported by Thuan \& Martin (1979) is for a 4\amine
beam at Arecibo centered on the galaxy.  The \HI\ mass given in Table 1
is consistent with their upper limits, particularly if the \HI\ is not
centered directly on the dSph, as is the case for several other
galaxies.

\subsubsection {And V}

This is an intriguing case not only because of the apparent association
of the \HI\ with the galaxy, but also because the \HI\ is listed as HVC
368 in the compilation of Wakker \& van Woerden (1991).  Thus a
measurement of the radial velocity of the galaxy that agrees with the
\HI\ velocity would provide the second firm association of an HVC with a
dSph (see the discussion of Sculptor below).  As shown in Figure 2,
however, HVC 368 lies quite close to HVC 287, also known as complex H.
The latter has an angular extent of $\sim$205 deg$^2$, and a velocity
similar to HVC 368.  Blitz et al.\ (1999) have argued that complex H
lies beyond the $\sim$40 kpc radius of the \HI\ disk of the Galaxy, but
probably not far beyond.  The proximity of the two clouds and the
similarity of their radial velocities suggests that the smaller cloud
may be a fragment of the larger one, and may thus be an unrelated
foreground object.  In that case, the probability of a chance
coincidence is much higher than the value of $\sim 10^{-2}$ for the
other dSph galaxies.  We present additional arguments in
\S 4 why this \HI\ cloud might be a chance superposition.  A measurement
of the optical velocity would clearly determine whether the \HI\ and the
galaxy are associated.

\begin{figure}[h!]
\figurenum{2}
\epsscale{1}
\plotone{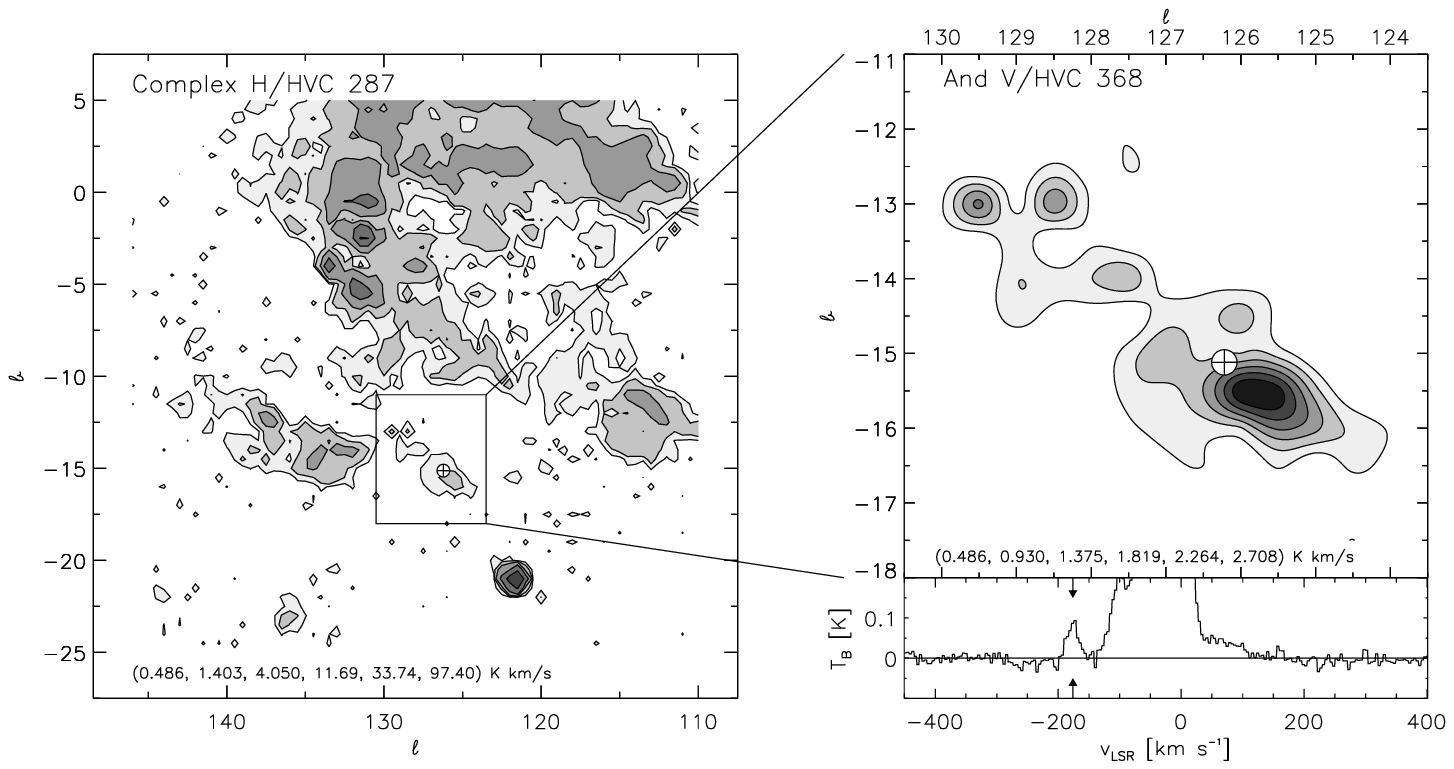}
\caption{\HI\ along the line of sight to And V.  It is unclear whether
HVC 368 is a fragment of HVC 287 or is a truly independent object.  The
blob of emission at \l = 122\deg, \b = -21\dege is \HI\ associated with
M31.}
\end{figure}

\subsubsection {DDO 210}

	The \HI\ in this galaxy was previously observed by Lo, Sargent, \&
Young (1993)
at the VLA, and the mass we obtain for the \HI\ emission centered on the galaxy
is in good agreement with theirs, correcting for the different distances
assumed.  Note, however the additional emission seen in Figure 1 that is
outside the Lo et al.\ (1993) field of view.
The relatively strong emission seen at \l~= 35\deg, \b~= -32\dege is a
narrow component at nearly the same velocity as the galaxy.  The low
level disconnected emission in Figure 1 may not be real, but the
\HI\ cloud associated with DDO 210 nevertheless 
appears to be considerably larger than that observed by Lo et al.\ (1993).

\subsubsection {Leo I}

There is a clear detection of an \HI\ cloud with a large velocity dispersion
in the direction of this galaxy at a velocity close to, but slightly
shifted from the velocity of the dSph.  The uncertainty in the velocity
of the line center is, however, relatively large and the width of the
line comfortably encompasses the velocity of the galaxy.  We show two
maps in Figure 3.  On the left is a map of the brightness temperature
distribution over the full width of the \HI\ line.  The map shows the
extent of what appears to be highly fragmented \HI\ emission; a larger
map (not shown) indicates that this emission is 
confined to the area shown in the panel.  As a
check, we produced a map over the same area within $\pm$10 \kms\ of
the Leo I optical velocity shown in the right hand panel of Figure 3.  
The \HI\ in this velocity range is confined to an area quite close to
the galaxy, but surprisingly, no \HI\ is seen directly
toward
the galaxy itself.  The full velocity extent of the \HI\ is much more
extended, and is seen over an area of about 10--20 deg$^2$.
Nevertheless, since the \HI\ in the velocity range of Leo I is so
closely confined to the environs of the galaxy,  it seems likely that
all of the \HI\ is at the distance of the dSph.  Higher resolution \HI\
mapping might make the association clearer.  The \HI\ non-detection
reported by Knapp, Kerr, \& Bowers (1978) was made with a single
pointing of the 300 ft (91 m) telescope at Green Bank and a 10\amine
beam; it is consistent with the present detection.

\begin{figure}[h!]
\figurenum{3}
\epsscale{1}
\plotone{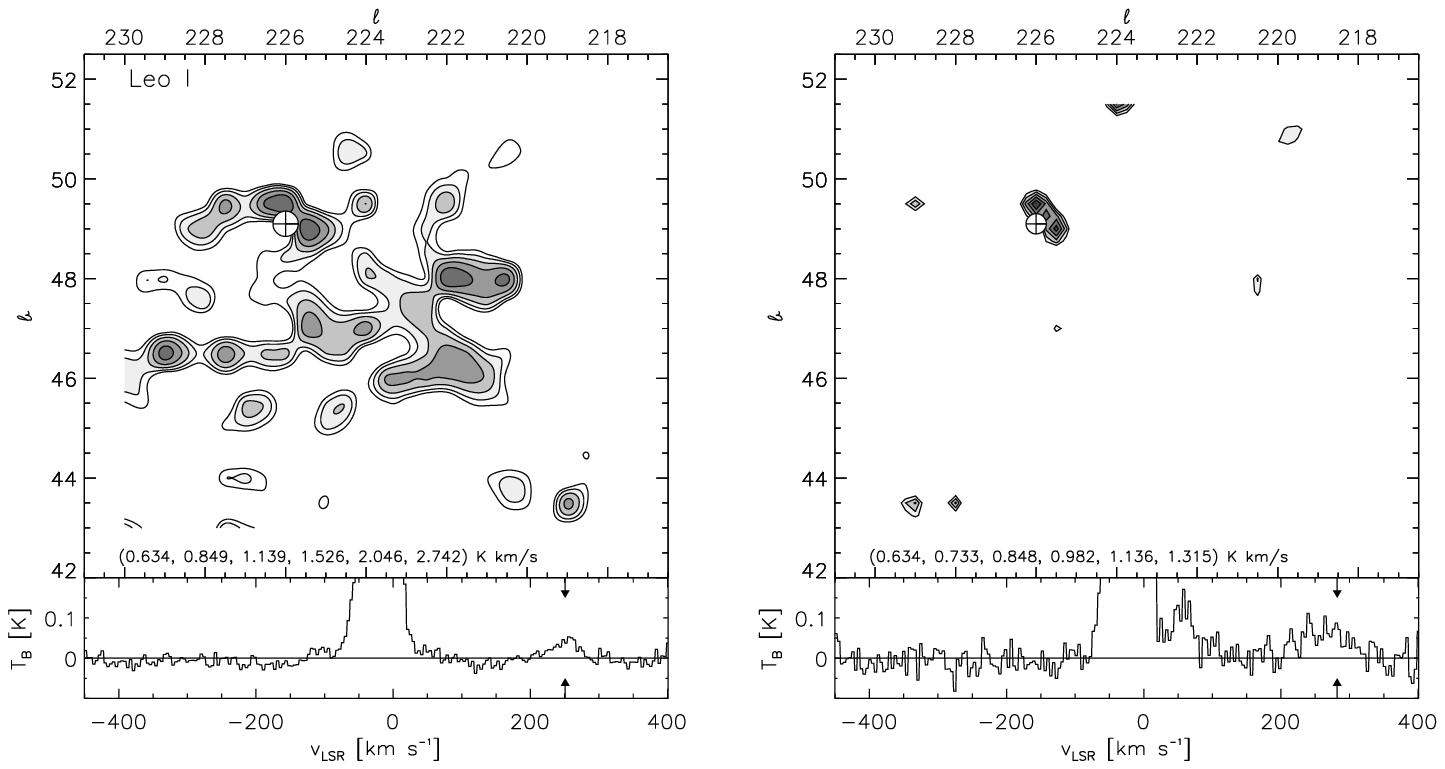}
\caption{Leo I.  ({\it Left}) Moment map for $\rm 237 \leq V_{LSR}
\leq 277$ \kms, covering the FWHM velocity of the Gaussian fit to the
position-averaged profile. ({\it Right}) Moment map for $\rm 271 \leq
V_{LSR} \leq 292$ \kms\ which covers the radial velocity of the galaxy
$\pm 10$ \kms.}
\end{figure}

\subsubsection {LGS 3}

The \HI\ cloud in the vicinity of LGS 3 was noted by Hulsbosch \& Wakker 
(1988) and is quite close to the galaxy, but the
velocity centroid of most of the \HI\ differs from that of the galaxy
by about 50 \kms.  This galaxy was mapped with the VLA by Young \& Lo
(1997) who found an \HI\ cloud centered on the dSph within a few \kms\
of the velocity of LGS 3 and a total \HI\ mass of $4 \times 10^5$
\Msun.  Smoothing of the LDS \HI\ profile at the position of the
galaxy shows both the -285 \kms\
component and a component centered at about -340 \kms.  A hint of the
more extreme velocity can be seen in the 140\amine spectrum shown by
Young \& Lo (1997).  The probability of two HVCs seen along the same
line of sight that are this compact is about 10$^{-2}$ (see
\S\ref{sec:analysis}).  While the large velocity difference between the cloud
shown in Figure 2 and the cloud mapped by Young \& Lo (1997) is
puzzling, the relation of the dSph to the cloud bears some resemblance
to the other systems pictured in Figure 1.  We discuss in \S 4 the possibility that the more negative velocity component may result from ram-pressure
stripping.

	Although the emission detected by Young \& Lo (1997) is not seen in
either the map or the spectrum shown in Figure 1, both are consistent
with their detection.  The peak flux density of this galaxy observed at
the 140\amine telescope is about 110 mJy, corresponding to an expected
antenna temperature of 0.021 K when observed with the Dwingeloo
telescope.  This is the strength at which the feature is seen, within
the noise, in the LDS at the position of the galaxy.  The spectrum
shown in Figure 1 is an average of 6 positions at which the galaxy is
detected; the expected peak temperature of that feature is lowered by yet
another factor of 2.5; a feature of that strength is too weak to be
detected.  Both the spectrum and the map of this source are
consistent with the observations of Young \& Lo (1997).

\subsubsection {Pegasus}

	Pegasus has been previously 
mapped by Lo, Sargent, \& Young (1993).  The mass
obtained with the LDS data is in reasonable agreement with that of Lo et
al.\ (1993) when account is taken of the different distances assumed and
the possibility that some of the flux may be missed with the VLA
observations.

\subsubsection {Phoenix}

Although this galaxy is too far south for the LDS, it has previously
been mapped by Carignan, Demers, \& C\^ot\'e (1991), Young \& Lo (1997) and 
St-Germain et al.\ (1999).  Because this galaxy lies close 
to the main emission from the Milky Way, and its optical velocity has
not yet been measured, it is unclear which of two relatively compact
velocity components, if either, is associated with the galaxy.  The \HI\
properties listed in Table 1 are from St-Germain et al.\ (1999).  They cite
evidence suggesting that the lower mass (i.e.\ $2 \times 10^5$ \Msun) \HI\
component is associated with the galaxy, in part because they find the
larger mass component implausibly large.  We follow their suggestion
and use the lower value in Figure 4 below.  Without a reliable single
dish map however, it is unclear whether the interferometer recovers all
of the flux associated with the galaxy.  Thus the mass quoted by
St-Germain et al.\ (1999) should be considered a lower limit.

\subsubsection {Sextans}

Although the emission from Sextans shown in Figure 1 is not centered
on the galaxy, the velocity of the \HI\ emission is quite close to the
optical velocity. The probability of a cloud outside the galaxy with a
velocity within $\pm$15 \kms\ of the galaxy is about 0.05.  The joint
probability of a small cloud within 4 square degrees with a
velocity outside of the normal Galactic emission in such close agreement 
with the optical velocity is about $1.5 \times 10^{-3}$.  The cloud
shown in Figure 1 is therefore probably associated with the
galaxy; the edge of 
the \HI\ cloud is only about 1 kpc from the galaxy in projection.

\subsubsection {Sculptor}

Sculptor is a particularly interesting case because of the two-lobed
\HI\ structure found by Carignan et al.\ (1998) and because the \HI\
emission associated with the galaxy is catalogued as HVC 561 in the
compilation of Wakker \& van Woerden (1991).  The map of the galaxy
shown in Figure 1 is incomplete because the galaxy is beyond the
southern declination limit of the LDS.  Carignan et al.\ (1998) were at
pains to point out that the \HI\ emission in their map might only be a
small fraction of the total because of the spatial filtering of their
interferometric observations, and because the \HI\ extends to the limit
of the spatial scale to which their observations are sensitive.   A
more complete map was recently published by Carignan (1999) using the
Parkes 43-m telescope showing that the \HI\ associated with Sculptor has
an extent of about 2\deg, larger than that shown in Figure 1,
but the total mass is not given.  The mass
given in Table 1 is from the LDS, but may be low by a factor of about 2-3.

\subsubsection {Tucana}

	Tucana is below the southern declination limit for the LDS, but
was observed at the ATNA by Oosterloo, Da Costa, \& Staveley-Smith
(1996), who found an \HI\ cloud close to, but not quite coincident with
the galaxy.  The centroid of the emission is located only 15\amine from
the galaxy and some of the emission is as close as 1\amine from it.
The joint probability of such close agreement in position and velocity
between the observed \HI\ and the optical galaxy is about 10$^{-4}$,
suggesting that the \HI\ cloud is indeed bound to the galaxy.
Furthermore the Oosterloo et al.\ (1996) observations were done with the
ATNA, and the map extends over a large fraction of the beam. It is therefore 
likely that the extent of the \HI\ emission is larger than that shown in their map, similar to what is observed in Sculptor.  The relationship of the
\HI\ to the optical galaxy in this case is similar to galaxies
such as Leo I, And V, Sculptor and possibly LGS 3.
The low probability of a chance superposition, 
together with the similarity to other systems, 
suggests that the \HI\ is indeed related to Tucana. If the cloud
is at the distance of the dSph, it has a mass of $1.5 \times 10^6$
\Msun.

\subsection {The Non-Detections}

A number of the non-detections are somewhat ambiguous because the
optical velocity of the galaxy is close to the Galactic \HI\
emission, or because a feature of marginal significance is present in
the LDS. High significance non-detections were made toward And II, Leo II and
And VII using the 140\amine telescope with an rms noise temperature of
about 35 mK in 1 \kms\ channels.  The non-detection toward Leo II is
consistent with that of Young (1999).  

A reported detection of Antlia by Fouqu\'e et al.\ (1990) could not be
reproduced, and is probably part of the emission
from NGC 3109 entering a sidelobe of the telescope.  We were unable to
see any evidence for emission in the LDS at the position observed by
these authors.

\setcounter{table}{1}

\begin{deluxetable}{lrrcc}
\tablecaption{Local Group Dwarf Spheroidals Undetected in the LDS}
\tablehead{
\colhead{Galaxy}   &
\colhead{\l} &
\colhead{\b} &
\colhead{dist [kpc]} &
\colhead{V$_{{\rm \scriptstyle{LSR}},opt}$ [\kms]}}
\startdata
And I   &      121.7   &  -24.9        &  \phn805\plm 40\phn   &  \nodata
\\
And II\tablenotemark{\alph{footnote}}\addtocounter{footnote}{1}  &      128.9   &   -29.2       &  \phn525\plm 110    &  -188\plm3
\\
And VI\tablenotemark{\alph{footnote}}\addtocounter{footnote}{1}  &
106.0 & -36.3    & \phn775\plm 35\phn  & \nodata                         \\
And VII\tablenotemark{\alph{footnote}}\addtocounter{footnote}{1} &              109.0  & -10.1          &  \nodata               & \nodata \\
Antlia	&	263.1	&	22.3	&	1235\plm65\phn	&	\nodata	\\
Leo II  &       220.2   &       67.2    &        \phn205\plm 12\phn     & \phantom{-}\phn77\plm 2
\\
Ursa Minor	&	105.0	&	44.8	&	\phn\phn66\plm3\phn\phn	&	-237\plm2	\\
Draco	&	86.4	&	34.7	&	\phn\phn82\plm6\phn\phn	&	-279\plm2 \\
Sagittarius	&	5.6	&	-14.1	&	\phn\phn24\plm2\phn\phn	&	\phantom{-}148\plm5	\\
KK99 191.1	&	109.0	&	-3.6	&	\nodata	&	\nodata	\\
KK99 348.1	&	109.1	&	-22.4	&	\nodata	&	\nodata \\
\sidehead{Local Group Dwarf Spheroidals Outside Observed Region of LDS}
Carina	&	260.1	&	-22.2	&	\phn101\plm5\phn\phn	&	\phantom{-}209\plm3	\\
Fornax	&	237.1	&	-65.7	&	\phn138\plm8\phn\phn	&	\phantom{-}\phn41\plm3	\\	
\enddata
\setcounter{footnote}{1}\tablenotetext{\alph{footnote}\addtocounter
{footnote}{1}}{Radial velocity from C\^ot\'e et al.\ (1999)}\tablenotetext{\alph{footnote}\addtocounter
{footnote}{1}}{\l, \b\ \& distance from
Armandroff, Davies, \& Jacoby (1998)}
\tablenotetext{\alph{footnote}\addtocounter
{footnote}{1}}{\l, \b\ from Karachentsev \&
Karachentseva (1999)} \end{deluxetable}

\section{Discussion}

	The detection of \HI\ associated with many dSph
galaxies suggests that rather than being gas-free systems,
perhaps half of them are in fact gas-rich galaxies, with \HI\ that is a
substantial fraction of the luminous mass, M$_{\rm L}$, and in some cases
exceeding it. If we take M$_{\rm L}$/L${\rm_V}$ = 1-2 in solar units, then more
than half of the galaxies in Table 1, about 25\% of all of the dSphs,
 have \HI\ masses in excess of 50\% of the
luminous mass of the galaxy.  Most previous observations have
concentrated on searching for \HI\ only at the central position of a
galaxy largely because the beam of the \HI\ observations was as large or
larger than the galaxy itself.  The observations presented here show
that sometimes the \HI\ avoids the nucleus of the galaxy and is seen
over a large area around it (e.g.\ Leo I).

With the exception of And V, there is little doubt that the \HI\ clouds
shown in the maps in Figure 1 are associated with the galaxies, even
though optical radial velocities are not available for most dSphs. As
discussed in \S 2, the probability of a chance superposition with an
\HI\ cloud is quite small, typically about 0.01, and if a velocity is
available, the probability is $< 10^{-3}$.

	If all of the galaxies shown in Table 1 have ionized hydrogen
masses equal to that of their atomic hydrogen, and if the \HII\ were
smoothly distributed out to the same radius as the \HI, the typical
emission measure except for Phoenix and Tucana would be 10$^{-3}$ --
10$^{-4}$ cm$^{-6}$ pc, more than an order of magnitude below what is
currently detectable.  Thus the dSphs might harbor substantial quantities 
of ionized gas in addition to what is shown in Figure 1.

 	We might reasonably ask why some of the dSph galaxies have
large, massive \HI\ envelopes (such as Leo I), why some have relatively wimpy
\HI\ envelopes (such as Sextans and Sculptor), and why some seem to be
devoid of \HI\ entirely (such as Ursa Minor and Draco).  To approach this
question, we have plotted the the \HI\ mass of each of the dSph and dIrr
galaxies in the Local Group as a function of the distance from the
nearest giant spiral, either M31 or the Milky Way (MW) in Figure 4a.
The data for the dIrr galaxies are taken from Mateo (1998).

\begin{figure}[h!]
\figurenum{4}
\epsscale{1}
\plotfiddle{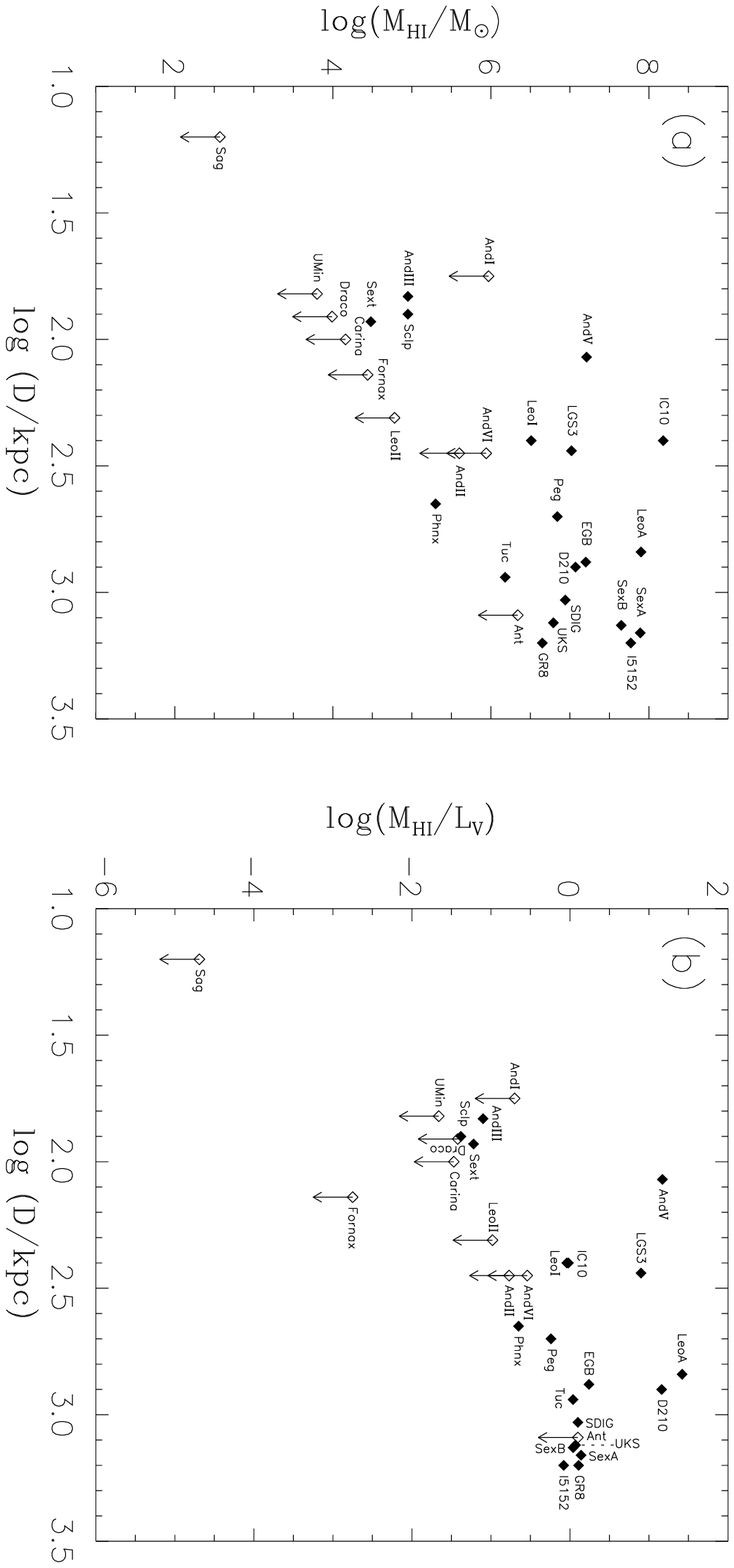}{170truemm}{180}{80}{80}{245}{560}
\vspace{.1in}
\caption{(a) Plot of the \HI\ masses of the Local Group dSphs and dIrrs 
vs.\ distance from the center of M31
or the Milky Way, whichever is closer.  (b) Same as (a) except that the
\HI\ mass is normalized by the the V-band luminosity in solar units.
}
\end{figure}

	Figure 4 shows a rather remarkable effect.  With the exception
of And V, no galaxy within 250 kpc of either M31 or the MW has
an \HI\ mass in excess of 10$^6$ \Msun; beyond 250 kpc almost all of the
dwarf galaxies have substantial \HI\ envelopes in excess of 10$^6$
\Msun, regardless of whether the galaxies are dSph or dIrr. Inside the
250 kpc cutoff most of the upper limits and the three detections are in
fact below 10$^5$ \Msun.  The one exception is And V which is the
galaxy most likely to be a chance superposition (see \S 3.1.2). 
Antlia and Phoenix are the exceptions beyond the 250 kpc 
cutoff. Antlia may be anomalous because of its close proximity to NGC 3109.
The \HI\ mass of Phoenix may be underestimated since its mass is
determined from 
an interferometric measurement confused by bright foreground
emission at the same velocity. It is also possible that another more
massive \HI\ component is actually associated with the galaxy
(\S4.1.7). 

	To check whether the effect seen in Figure 4a 
might be a distance effect related
to the Malmquist bias, we replotted the figure by normalizing to the visual
luminosity of each galaxy, L$_{\rm V}$, taken from Mateo (1998); the
results are shown in Figure 4b.  Evidence for the 250 kpc \HI\ cutoff
remains quite strong in this plot.  
Van den Bergh (1999a) noted that the
dSph galaxies tend to be closer to M31 and the Milky Way than the dIrr
galaxies, but because some of the galaxies beyond the 250 kpc \HI\ cutoff are
dSphs, the segregation in \HI\ properties is not simply a function of
morphological type.  The sharp cutoff implies that some process strips
the galaxies with perigalacticons $<$ 250 kpc of their \HI, a
suggestion first made by Einasto et al.\ (1974) and later by 
Lin \& Faber (1983).  We
investigate the relative importance of ram-pressure stripping and tidal
stripping below.

\subsection {Ram-Pressure Stripping}

	If the sharp boundary at 250 kpc is due to ram-pressure stripping
by hot halo gas (Gunn \& Gott 1972), one can estimate the density of
the ambient gas responsible for the stripping, $\rho_{\circ}$, from
\begin{equation}
\rho_{\circ}~~>~~ \alpha G \Sigma_* \Sigma_g/ V^2, 
\end{equation}
where $\Sigma_*$ and $\Sigma_g$ are the the stellar and gas
surface densities respectively in the dwarf galaxy, $V$ is the velocity of
the galaxy through the hot halo gas, and $\alpha$ is a constant near
unity that depends on whether the galaxy is a flattened or a spheroidal
system, and on the functional form of the gravitational potential.
For a uniform spherical stellar
system with an extended uniform density halo, $\alpha = \pi/6$. 
$\Sigma_*$ and $\Sigma_g$ in this case are the peak values measured at
the center of the galaxy. The inequality exists because the gas may be
clumped and the surface filling fraction of the \HI\ may be less than
unity, raising the local value of $\Sigma_g$ above its beam averaged
value.  If the gravitational mass of the galaxy,
M$_*$, is given by $Rv_{\rm3D}^2/G$, then 
we may rewrite Equation 1 as:
\begin{equation}
\rho_{\circ}~~>~~{{\Sigma_g v_{\rm3D}^2}\over{4 R V^2}}.
\end{equation}
\noindent
In terms of observables, Equation 2 may be rewritten as
\begin{equation}
n_{\circ}~~ >~~ 1.12 \times 10^{-4}~~ {{T_B (\Delta V)^3}\over{R V^2}}\ \, \rm cm^{-3},
\end{equation}
\noindent
where $n_{\circ}$ is the number density of the hot halo gas, $T_B$ is
the peak brightness temperature of the \HI, and $\Delta V$ is the measured
full width at half maximum of the \HI\ line.  We use this equation to
estimate $n_{\circ}$.  For $V$ we take the one-dimensional velocity
dispersion of 60 \kms\ for the Local Group dwarf galaxies (van den
Bergh 1999a).  For $\Delta V$, we take the mean value for the galaxies
from Table 1 of 25 \kms.  For $R$, we take it equal to $R_g$, the
radius to which gas is observed in a dwarf galaxy; it ranges from about
1 - 10 kpc for all of the galaxies in the sample with the larger values
all outside the 250 kpc cutoff radius.  Inside the cutoff radius,
three galaxies, And III, Sculptor and Sextans, still have associated \HI, and
we take $R = 2$ kpc, the largest of the radii to which gas
is observed. Thus $n_{\circ} \simgt 2.4 \times 10^{-5}$ \cc.
Interestingly, this value is close to the value of $\sim 5 \times
10^{-5}$ \cc\ derived by Moore \& Davis (1994) for ram-pressure
stripping of the Magellanic Clouds.  Their value applies to the
density of a hot halo at a distance of 65 kpc from the Galactic
Center, whereas ours is an average over a volume of 250 kpc radius. 

If the hot gas is in virial
equilibrium and the mass of the MW is 1 -- $1.5 \times 10^{12}$ \Msun, 
the hot gas temperature is $\sim$ 1 -- $1.4 \times 10^6$ K, and the
thermal pressure $P/k$ = 23 - 34 K \cc.  The total mass of the hot halo
at the derived $n_{\circ}$ is about $1 \times 10^{10}$ \Msun\ to 250
kpc.

	Gas will be stripped from the galaxy if the column density of
hot gas is equal to the column density in the galaxy.  Thus 
for galaxies (or HVCs) with N(\HI) $> n_{\circ} \times$ 250 kpc = $1.8
\times 10^{19}$ \cm2, the gas can remain in the galaxy until it
collides with the MW.  The most recent map of Sculptor by Carignan
(1999) shows \HI\ contours ranging from 0.15 - $2.4 \times 10^{19}$
\cm2, but with a broad emission plateau of about $0.6 \times 10^{19}$
\cm2, somewhat below the ram-pressure limit. The discrepancy, while not
large, may be due to clumping (implying higher mean column densities
within the beam), a somewhat high estimate for $n_{\circ}$, 
Sextans not having traversed a full 250 kpc in its
orbit through the hot halo, or some combination of the three.  The
radial velocity of Sculptor relative to the Galactic Standard of Rest
(GSR) is 75 \kms, close to the value of 60 \kms\ assumed.

\subsection{Tidal Stripping}

Gas will be tidally stripped from a galaxy if 
\begin{equation}
{{GM}\over {R^2}}~~>~~-{\frac {\rm d}{{\rm d}r}}\left({{\Theta ^2}\over r}\right)R,
\end{equation}
\noindent
where $r$ is the distance from the MW or M31 to the galaxy, 
and $\Theta$ is the circular
speed of the MW or M31 at the distance of the dwarf galaxy.
At a distance of 250 kpc, $\Theta$ = 130 \kms\ (appropriate for an
enclosed MW mass of $1 \times 10^{12}$ \Msun); the tidal radius of a
dwarf galaxy with a total mass of $1 \times 10^8$
\Msun\ is about 11.5 kpc, greater than the largest radius to which \HI\
is detected in any Local Group dwarf.  Thus for these parameters, the gas is
tidally stable.  For dwarf galaxies at 80 kpc from the MW, 
the distance of Sculptor and Sextans, the tidal radius is about 4 kpc,
comfortably larger than the 1.5 -- 2 kpc to which \HI\ is detected
in either Sculptor or Sextans.  Thus, the gas in these galaxies is
tidally stable at their current distance, but if their perigalacticon
is significantly smaller, then even the present day observed gas could
be tidally stripped from these galaxies.  Thus tidal stripping can be
important in removing the outer \HI\ envelopes of the dSph galaxies
within 250 kpc of the MW and M31.

Nevertheless, most of the dSphs within 250 kpc are devoid of \HI\ to the
current limits of detectability.  While Galactic tides can be important
in stripping the outermost \HI\ layers of Local Group dSphs, it is
difficult to understand how tidal stripping can completely rid a dSph
of its atomic gas.

It is instructive to compare the effect of tidal vs. ram-pressure
stripping in a dwarf galaxy at a radius of 10 kpc and located at a distance of 250 kpc from the MW.
The tidal acceleration from Equation 4 above is $R\Theta^2/r^2$. The
acceleration of a parcel of gas due to ram-pressure is 
\begin{equation}
{{\rho_{\circ} V^2}\over{\Sigma_g}}~~ =~~ {{\rho_{\circ} V^2}\over{{\rm N(\HI)} \mu
m_{\rm H}}}, 
\end{equation}
\noindent
where $\mu$ is the mean mass per nucleon of the atomic gas.  For
$\Theta = 130$ \kms, $n_{\circ}$ = $2.5 \times 10^{-5}$ \cc,
and N(\HI) = $5 \times 10^{18}$ \cm2,  a typical
value in the outskirts of Leo I and LGS 3, the ratio of the
ram-pressure acceleration to the tidal acceleration is about 18. 
Tidal stripping becomes relatively more important, however, as one gets
closer to the center of the Local Group giant spirals, 
but if there is
even one tenth the density of hot gas around the MW and M31, as we
derive above, ram-pressure stripping will be relatively 
more important than tidal stripping unless perigalacticon is very
small.

	We therefore conclude that although tidal stripping may play a
role in the gas depletion of the dwarf galaxies in the Local Group,
tidal stripping alone cannot explain the near total absence of
atomic gas in most of the dSphs within 250 kpc of either the MW or M31.
Ram-pressure stripping is much more efficient over most of the
parameter space permitted by the \HI\ observations, and implies that the
MW and M31 have hot halos with radii of $\sim$250 kpc, beyond which stripping
is ineffective.

	It is quite reasonable that the MW and M31 have hot halos 
in the context of the Blitz et al.\ (1999) picture of the formation of
the Milky Way and M31 from the accretion of HVCs.  The dynamical model
presented by these authors is quite simple and does not include
collisions between HVCs which are expected close to either galaxy or in
the region between the galaxies.  Such collisions are, however,
expected; the typical center-of-mass collision velocity is about 100
\kms\ at the present epoch, high enough to completely ionize the
colliding clouds and to raise their temperatures to $\sim 10^5$ K
(McKee \& Hollenbach 1980).  Collisions between clouds at the present
epoch are, however, expected to be rare, but should have been more
frequent in the early universe (see Fig.\ 16 in Blitz et al.\ 1999).  The
typical cloud collision velocity at those times would have been higher,
leading to hot gas temperatures which can reach the virial values.  The
cooling time of the hot halo gas at the present epoch is $\sim 3 \times
10^{10}$ yr at an inferred mean density of $2.5 \times 10^{-5}$ \cc.
Thus the hot halo gas is stable for more than a Hubble time,
but the inner parts can cool and condense in a shorter time if the
density distribution is isothermal.  Whether HVC collisions are
frequent enough to produce a hot halo can be tested by direct numerical
simulations.  The halos around each galaxy need not be separate
entities and may be connected along the line between the two galaxies.

	The existence of a hot halo is consistent with the 
destruction of $\sim$1000 objects with \HI\ content of $\sim 10^7$ \Msun,
close to the typical mass derived by Blitz et al.\ (1999) for the HVCs.
It is worth noting that Klypin et al.\ (1999) have shown that
simulations of hierarchical structure formation seem to require about 1000 dwarf galaxies to have formed in the Local Group, one and a
half orders of magnitude more than have currently been identified.
Klypin et al.\ (1999) suggested that the required number of dwarfs might be
consistent with the inferred population of HVCs, but it is also
possible that the gaseous halo is from a population of true dwarf
galaxies with a typical \HI\ content similar to that found in the present-day dwarfs that have been destroyed by collisions. A mass of $10^{10}$
\Msun\ of hot gas would be provided by about $10^2 - 10^4$ dwarf
galaxies, with typical \HI\ masses between $10^6 - 10^8$, the range of
\HI\ masses in present-day dwarfs not yet stripped of \HI. 
The mean free path for collisions between HVCs ought to be smaller than
that for dwarfs because of their larger diameters, and the difference
between the two cases might be testable by simulations.

\subsection{The Velocity Anomalies of LGS 3 and Leo I}
	
One of the more difficult of the \HI\ observations to understand is that
of LGS 3, and to a lesser extent, Leo I.  LGS 3 has an \HI\ cloud clearly
associated with it, but the \HI\ cloud differs in velocity from the
galaxy by 50 \kms.  The observations of this galaxy by Young \& Lo
(1997) show a well-defined \HI\ component centered on the galaxy with a
velocity essentially identical to the systemic velocity of the stars.
The extent of the gas centered on the galaxy is only about 5\arcmin\
compared to the extent of about 1\degper5 in Figure 1.  With a
velocity difference of 50 \kms, the gas shown in Figure 1 cannot be
gravitationally bound to LGS 3, which suggests a chance superposition.
On the other hand, the probability of a chance spatial superposition is about
1\%, and the probability that the velocity would be so close to the
systemic velocity is about 0.2, for a joint probability of about $2
\times 10^{-3}$.

The situation with Leo I is similar but less extreme.  The right hand
panel of Figure 3 shows that \HI\ at the velocity of the galaxy is
closely associated with it, even though no \HI\ is detected 
toward the stars themselves.  Nevertheless, the velocity centroid of
the emission shown in the right hand panel of Figure 3 differs from the 
systemic velocity of the galaxy by about 30 \kms.  In this case,
however, the gas shown in both panels is part of the same velocity
component.  While the extended emission shown in Figure 3 is unlikely
to be gravitationally bound to Leo I, it is more difficult to argue that
this gas is also a chance superposition, not only because the lower
velocity gas is part of the same velocity component, but also because
the joint probability of having two galaxies in this small sample with
two \HI\ clouds in the same line of sight at velocities close to the
systemic velocity is $<$ 10$^{-5}$.

We note, however, that both galaxies are close to the 250 kpc 
boundary shown in Figure 4.  Leo I is 250 kpc from the Milky Way (Mateo
1998) and LGS 3 is 270 kpc from the center of M31. Could it be that the
\HI\ seen in Figure 1 for each galaxy is beginning to be stripped by the
ram-pressure of the hot halo gas?  Certainly for LGS 3, the gas in
Figure 1 looks as if it has been swept away from the galaxy itself.  Tidal
stripping is not an option for either galaxy because there is a
systematic offset in velocity in only one sense; tidal stripping would
produce plumes with velocities both larger and smaller than the
systemic velocity of the dSph.  From Equation 5, we find that the
mean acceleration of the gas due to tidal stripping is $\sim 1.3 
\times 10^{-10}$ cm s$^{-2}$.  Gas is detected as much as 1\degper5
from LGS 3, or about 20 kpc in projection 
from the galaxy. With a velocity difference
of 50 \kms, gas farthest from the galaxy will have taken $4
\times 10^8$ yr to reach that distance.  Assuming that the acceleration
is constant, ram-pressure stripped gas would attain a velocity of
$\sim$15 \kms\ in that time. This is a bit on the low side, but the
agreement is not unreasonable, given the uncertainty with which the
critical parameters are known.  The density of the hot gas might be
somewhat higher, as might the velocity of the galaxy with respect to
the hot halo gas, and projection effects can be important at the 50\%
level.  

If the position and velocity offsets of the \HI\ from LGS 3 are due
to ram-pressure stripping,  one would expect there to be a velocity gradient in
the sense that the most extreme differences in velocity are seen
farthest from the galaxy.  The individual \HI\ spectra that compose
Figure 1 do not, however, show any detectable gradient in the
extended cloud. 

Ram-pressure stripping should produce velocity differences from the
systemic velocity with well-determined signs: the velocity of the
stripped gas should always be closer to the systemic velocity of either
the MW or M31 than that of the dwarf being stripped.  In the case of Leo I, the
galaxy has a GSR velocity of +178 \kms; the anomalous velocity gas has
a GSR velocity of +158 \kms, in accord with expectations.
For LGS 3, the velocity relative to
the GSR is about 25 \kms\ more negative than that of M31, suggesting that
the stripped gas should have a more positive velocity.  But this is
not the case; the stripped gas is 50 \kms\ more negative.  This 
discrepancy is difficult to reconcile with ram-pressure stripping.

We conclude that ram-pressure stripping can plausibly produce the
magnitude of the velocity and positional offsets for LGS 3 and Leo I if
the velocity of the galaxy relative to the hot gas and the density of
the hot gas are within factors of two and three respectively of the
values assumed for them.  The sign of the velocity offset for LGS 3,
however, seems inconsistent with ram-pressure stripping.

\subsection{Comparison of Dwarf Galaxy Envelopes with HVC Properties}

	In their paper on HVCs, Blitz et al.\ (1999) derived 
properties of HVCs under the assumption that 
they are typically at a distance of 1 Mpc; however
they did not correct for beam smearing, which would lower both the
diameters and the derived masses. Furthermore, if the typical HVC has a
distance more like that of M31 than the 1 Mpc assumed, 
their derived masses may be too high by as much
as a factor of 4, and their diameters too high by a factor of 2.  
A complete northern
hemisphere HVC catalogue by Robishaw \& Blitz (in preparation) will
correct for beam smearing. Given this range of uncertainty,
typical HVC \HI\ masses are about 5 - $20 \times 10^6$ \Msun\ if the HVCs are
extragalactic and typical diameters are
about 15 - 28 kpc.  Some of the dwarf galaxy \HI\ envelopes are just in
this range, notably DDO 210, Leo I, LGS 3, IC 10, Leo A, Sextans A, and
possibly And V.  Several others are confused with the velocities from
Galactic foreground emission, and although their masses are in the
correct range, the full extent of their \HI\ emission is poorly
determined.  Thus, numerous Local Group dwarfs have \HI\ properties
virtually indistinguishable from extragalactic HVCs, and as pointed 
out in \S4, two previously catalogued HVCs are apparently associated
with galaxies.

	This suggests that some of the HVCs might harbor undetected low
surface brightness (LSB) galaxies, and searches are currently underway
to detect galaxies toward the HVCs.  If successful, this would bridge
the gap between the non-detections of HVC analogues without stars
in deep extragalactic \HI\ searches (Zwaan et al.\ 1997; Spitzak \&
Schneider 1998).  On the other hand, the extragalactic 
\HI\ searches have sensitivities
that trail off just at or above the derived typical HVC mass,
especially if the masses derived by Robishaw \& Blitz (in preparation)
turn out to be lower, as expected. So there may nevertheless be a
substantial population of low-mass intergalactic \HI\ clouds without
associated stars.   In either case, the HVCs might then
turn out to be the missing dwarf galaxies in the simulations of Klypin
et al.\ (1999).

\subsection{Implications for Galaxy Formation and Evolution}
\label{sec: implications}

	Mateo (1998), Grebel (1999) and van den Bergh (1999b) discuss 
several problems associated with the
apparent lack of interstellar gas in the dSph galaxies, most notably,
their complex star formation histories.  Some galaxies seem to have had
several episodes of star formation, including some as recently as 1 --
2 Gyr ago (Grebel 1999), but this hardly seems possible without at least
some traces of gas that could have fueled this activity.  The \HI\
observations presented in this paper suggest that one way around this
problem is that {\it all} of the Local Group dwarf galaxies have had
loosely bound \HI\ envelopes such as those seen for the galaxies at
distances beyond the 250 kpc cutoff radius when the last episode of
star formation took place.  This loosely bound gas would be subject to
small relatively localized perturbations that could lead either to star
formation or gas disruption and may be why the star formation
histories of the Local Group dwarfs are so heterogeneous (Mateo 1998;
Grebel 1999). Dissipation in the gas might have generated both low
levels of ongoing star formation as well as occasional large bursts
until the orbits of the galaxies brought them within the hot
halos around the MW and M31.  Star formation would then have ended for
the galaxies with \HI\ column densities insufficient to withstand the
ram-pressure stripping.  Clearly, the extended
\HI\ envelope around Leo I is plausibly the source of the relatively
recent star formation activity in that galaxy (e.g.\ Grebel 1998).

It is perhaps only when the galaxies venture within a
radius of 250 kpc that the dwarfs become stripped of their \HI,
eventually losing their ability to form stars.  The dynamical
simulations of HVCs in the context of the evolution of the Local Group
(Blitz et al.\ 1999) suggest that at least some of the dwarf galaxies
inside the 250 kpc cutoff may be approaching the MW and M31 for the
first time.  To traverse the entire length of the hot halo for galaxies
at a velocity of 60 \kms\ appropriate for the Local Group dwarfs (van den Bergh 1999a) takes $8 \times 10^9$ yr, a substantial fraction of a Hubble time. Thus some of the dwarfs such as Carina may only recently have lost
their gas, while others such as Ursa Minor, that show no
evidence of recent star formation activity, may have orbits that kept
them within the cutoff distance for most of a Hubble time.  

If the  connection between LSB dSph
galaxies and the HVCs can be confirmed, it might also solve the
metallicity problem for HVCs. That is, if the
HVCs are extragalactic, they must have existed for a Hubble time, but
the abundances and metallicities, while low, are not primordial (Wakker
\& van Woerden 1997; Wakker et al.\ 1999).  How then did these HVCs get
their metals?  If  the HVCs are associated with low levels of star
formation as in the dSph galaxies, then the stars themselves could have
contaminated the gas.  With a range of [Fe/H] $\simeq$ -1 to -2 dex for
the stars in
dwarf galaxies, one expects the gas to reflect these values.
The best determined metallicity toward an
HVC (complex C) is the measurement of S/H = 0.09 times the solar
value (Wakker et al.\ 1999), as expected, but more measurements are
needed.

\section{Summary}
\label{sec:summary}

	We have shown that there is good evidence that nearly half
of the dwarf spheroidal galaxies in the Local Group
contain large quantities of gas in an extended distribution around each
galaxy.  Even without measured velocities for most of the galaxies,
the associations of the gas with the galaxies must be real because the
number of positional coincidences is more than two orders of
magnitude greater than would be expected from random placements of both
the galaxies and the HVCs.  The properties of the \HI\ associated with
many of the dwarfs are similar to those expected for extragalactic
HVCs, suggesting that the HVCs may harbor LSB galaxies similar to the
dwarf galaxies in the Local Group.

	We have investigated the reason for the great diversity in the
neutral gas content of Local Group dwarf galaxies, and have shown
that within 250 kpc of the center of both the MW and M31, the gas
content drops precipitously.  Both tidal and ram-pressure stripping
can play a role in removing the gas, but ram-pressure stripping is more
effective and can strip a galaxy completely. The inferred mean density of
hot gas is $\sim 2.5 \times 10^{-5}$ \cc. 

	Two of the Local Group dwarfs, LGS 3 and Leo I, have \HI\ envelopes
that differ from the systemic velocities by 50 \kms\ and 30 \kms\
respectively.  The joint probability that both galaxies have unrelated
clouds along the line of sight is $\la$ 10$^{-5}$.  
Ram-pressure stripping could cause large velocity offsets, and
in the case of Leo I, the inferred hot halo properties are in
reasonable quantitative agreement with what is observed.  For LGS 3,
the agreement is less good, but may still be within the range of
acceptability.

	We find that the \HI\ observations can explain qualitatively
the diversity in the star formation histories of the Local Group
dwarfs, both the presence and absence of recent star formation in 
individual dSphs.  

\acknowledgments

	We are grateful to Jay Lockman for obtaining \HI\ spectra at
the 140\arcmin\ telescope to help confirm some of the Leiden-Dwingeloo
results and Tom Oosterloo for providing data on the Tucana dwarf.  
We would like to thank Taft Armandroff, Julianne Dalcanton, 
Eva Grebel, Raja Guhatakurtha, Dave Hollenbach, Chris McKee, David
Spergel, Hy Spinrad, Amiel Sternberg, and Sidney van den Bergh for
useful and lively discussions, some of which have been conducted
electronically.  The referee, Mario Mateo, made numerous useful
suggestions, and provided the velocity of And III prior to publication.




\begin{references}

\reference {} Armandroff, T.E.,  Jacoby, G.H., \& Davies, J.E. 1999, AJ, 118,
1220 

\reference {} Armandroff, T.E., Davies, J.E., \& Jacoby, G.H. 1998, AJ,
116, 2287

\reference {} Banks, G.D., et al.\ 1999, ApJ, 524, 612

\reference {} Blitz, L., Spergel, D.N., Teuben, P.J., Hartmann, D., \& Burton,
W.B. 1999, ApJ, 514, 818

\reference {} Braun, R., \& Burton, W.B., 1999, A\&A, 341, 437

\reference {} Carignan, C. 1999, PASA, 16, 18

\reference {} Carignan, C., Beaulieu, S., C\^ot\'e, S., Demers, S., \&
Mateo, M. 1998, AJ, 116, 1690

\reference {} Carignan, C., Demers, S., \& C\^ot\'e, S. 1991, ApJ,
381, L13

\reference {} C\^ot\'e, P., Mateo, M., Olszewski, E.W., \& Cook, K.H.
1999, ApJ, 526, 147

\reference {} C\^ot\'e, P., Mateo, M., \& Sargent, W.L.W. 2000, in preparation.

\reference {} Einasto, J., Saar, E., Kaasik, A., \& Chernin, A.D. 1974, Nature, 252,
111

\reference {} Fouqu\'e, P., Durand, N., Bottinelli, L., Gouguenheim, L.,
\& Paturel, G. 1990, A\&AS, 86, 473

\reference {} Gallart, C.,  Mart\'{\i}nez-Delgado, D.,  Aparicio, A., \&
Freedman, W.L. 1999, in {\it The Stellar Content of
Local Group Galaxies}, Whitelock \& Cannon, eds., ASP:San Francisco, p. 284

\reference {} Grebel, E. 1999, in {\it The Stellar Content of Local Group
Galaxies}, Whitelock \& Cannon, eds., ASP:San Francisco, p. 17

\reference {} Gunn, J.E., \& Gott, J.R. 1972, ApJ, 176, 1

\reference {} Hartmann, D., \& Burton, W.B. 1997, Atlas of
Galactic Neutral Hydrogen (Cambridge University Press:Cambridge)
(LDS)

\reference {} Hulsbosch, A.N.M., \& Wakker B.P. 1988, \aaps,  75, 191

\reference {} Karachentsev, I.D., \& Karachentseva, V.E. 1999, \aa 341, 355

\reference {} Klypin, A.A., Kravtsov, A.V., \& Valenzuela, O. 1999, ApJ,
in press

\reference {} Knapp, G.R., Kerr, F.J., \& Bowers, P.F. 1978, AJ, 83, 360

\reference {} Lin, D.N.C., \& Faber, S.M. 1983, ApJ, 266, L21

\reference {} Lo, K.Y., Sargent, W.L.W., \& Young, K. 1993, AJ, 106, 507

\reference {} Mateo, M. 1998, ARAA, 36, 435

\reference {} McKee, C.F., \& Hollenbach, D.J. 1980, ARAA, 18, 219

\reference {} Moore, B., \& Davis, M. 1994, MNRAS, 270, 209

\reference {} Oosterloo, T., Da Costa, G.S., \& Stavely-Smith, L., 1996,
AJ, 112, 1969

\reference {} Sembach, K.R., Savage, B.D., Lu, L., \& Murphy, E.M. 1999, ApJ, 515, 108

\reference {} Spitzak J.G., \& Schneider, S.E. 1998, ApJS, 119, 159

\reference {} St-Germain, J., Carignan, C., C\^ot\'e, S., \& Oosterloo, T.
1999, AJ, 118, 1235 

\reference {} Thuan, T.X., and Martin, G.E. 1979, ApJ, 232, L11

\reference {} van den Bergh, S. 1999a, in {\it The Stellar Content of
Local Group Galaxies}, Whitelock \& Cannon, eds., ASP:San Francisco, p. 3

\reference {} van den Bergh, S. 1999b, A\&A Reviews, 9, 273 

\reference {} Wakker, B.P., \& van Woerden, H. 1991, A\&A, 250, 509

\reference {} Wakker, B.P., \& van Woerden, H. 1997, ARAA, 35, 217

\reference {} Wakker, B.P., et al.\ 1999, Nature, 402, 388

\reference {} Weiner, B., Vogel, S., \& Weymann, R. 2000, ApJ, in preparation.

\reference {} Whiting, A.B., Irwin, M.J., \& Hau, G.K.T. 1997, AJ, 114, 996

\reference {} Young, L.M., \& Lo, K. Y. 1997, ApJ, 490, 710

\reference {} Young, L.M. 1999, AJ, 117, 1758

\reference {} Zwaan, M., Briggs, F.H., \& Sprayberry, D. 1997, ApJ,
490, 173 

\end{references}
\end{document}